\documentstyle[aps,epsf,rotate,eqsecnum,multicol]{revtex}
\title{Kinetic theory of shot noise in non-degenerate diffusive conductors}
\author{H. Schomerus,$^1$
E. G. Mishchenko,$^{1,2}$
 and C. W. J. Beenakker$^1$}
\address{{}$^1$Instituut-Lorentz, Universiteit Leiden,
P.\,O.~Box 9506, 2300 RA Leiden,
  The Netherlands
\\
{}$^2$L. D. Landau Institute for Theoretical Physics, Kosygin 2, Moscow
117334, Russia
}
\newcommand{\bcols}{\ifpreprintsty\else\begin{multicols}{2}\fi}
\newcommand{\ecols}{\ifpreprintsty\else\end{multicols}\fi}
\begin{document}
\draft
\date{January 1998}
\maketitle

\begin{abstract}
We investigate current fluctuations in non-degenerate semiconductors,
on length scales intermediate between the elastic and inelastic mean
free paths.  We present an exact solution of the non-linear kinetic
equations in the regime of space-charge limited conduction, without
resorting to the drift approximation of previous work. By including
the effects of a finite voltage and carrier density in the contact
region, a quantitative agreement is obtained with Monte Carlo simulations
by Gonz{\'a}lez {\it et al.}, for a model of an energy-independent
elastic scattering rate. The shot-noise power $P$ is suppressed below
the Poisson value $P_{\rm Poisson}=2e\bar I$ (at mean current $\bar I$)
by the Coulomb repulsion of the carriers. The exact suppression factor
is close to $1/3$ in a three-dimensional system, in agreement with the
simulations and with the drift approximation. Including an energy
dependence of the scattering rate has a small effect on the suppression
factor for the case of short-range scattering by uncharged impurities or
quasi-elastic scattering by acoustic phonons. Long-range scattering by
charged impurities remains an open problem.
\end{abstract}

\bcols
\section{Introduction}

The kinetic theory of non-equilibrium fluctuations in an electron gas was
pioneered by Kadomtsev in 1957 \cite{Kadomtsev} and fully developed ten
years later \cite{Kogan,Gantsevich}. The theory has been
comprehensively reviewed by Kogan \cite{Book}.
In recent years there has been a revival of the interest in this field, because
of the discovery of fundamental effects on the mesoscopic length scale.
(See Ref.\ \cite{reviews} for a recent review.) One of these effects is
the sub-Poissonian shot noise in degenerate electron gases
on length scales intermediate between the
mean free path $\ell$ for elastic impurity scattering and the inelastic
mean free path $\ell_{\rm in}$ for electron-phonon or electron-electron
scattering. The universal one-third suppression of the shot-noise power
predicted theoretically \cite{Buettiker,Nagaev} has been observed
in  experiments on semiconductor or metal wires of micrometer length
\cite{exp1,exp2,exp3,exp4}.

The electron density in these experiments is sufficiently high
that the electron gas is degenerate. The reduction of the shot-noise
power
\begin{equation}
\label{eq:powerdef}
P=2\int_{-\infty}^\infty{\rm d}t'\,\overline{\delta I(t) \delta I(t+t')}
\end{equation}
[with $\delta I(t)$ the fluctuations of the current around the mean
current $\bar I$]
below the Poisson value 
$P_{\rm Poisson}=2e\bar I$ is then the
result of correlations induced by the Pauli exclusion principle.
When the electron density is reduced, the Pauli principle becomes ineffective.
One enters then the regime of a non-degenerate electron gas, studied 
recently in Monte Carlo simulations by Gonz{\'a}lez {\em et al.}
\cite{Gonzalez}. In a model of energy-independent three-dimensional
elastic impurity scattering these authors found the very same ratio
$P/P_{\rm Poisson}=1/3$ as in the degenerate case. The origin of the
suppression is quite different, however, being due to correlations 
induced by long-range
Coulomb repulsion --- rather than by the Pauli principle.
The one-third suppression of shot-noise in the computer simulations
required a large voltage and short screening
length, but was found to be otherwise independent of material
parameters. 

Subsequent analytical work by one of the authors \cite{Beenakker1998}
explained this universality as a feature of the regime of space-charge
limited conduction. 
The kinetic equations in this regime are highly
non-linear, and could only be solved in the approximation that the
diffusion term is neglected compared to the drift term.
This is a questionable approximation: The ratio of the two terms is
$1/d$, with $d$ the dimensionality of the density of states.
The result of Ref.\ \cite{Beenakker1998},
\begin{equation}
\label{eq:analyt}
P/P_{\rm Poisson}=\frac {12} 5 \frac{3d^2+22d+64}{(d+2)(3d+4)(3d+8)}
,
\end{equation}
becomes exact in the large-$d$ limit, when 
$P/P_{\rm Poisson}\to 4/5d$, but has an error of unknown
magnitude for the physically relevant value $d=3$.

The main purpose of the present paper is to report the exact solution of
the kinetic equations in the space-charge limited transport regime.
We find that inclusion of the diffusion term has a pronounced effect on
the spatial dependence of the electric field, although the ultimate
effect on the noise power turns out to be relatively small:
The exact suppression factor differs from Eq.\ (\ref{eq:analyt}) by
about $10\%$ for $d=3$. 
We find
\begin{equation}
\label{eq:pownum}
P/P_{\rm Poisson}=\left\{
\begin{array}{cc}
0.6857& \quad\mbox{for}\quad d=1,\\
0.4440& \quad\mbox{for}\quad d=2,\\
0.3097& \quad\mbox{for}\quad d=3,
\end{array}
\right. 
\end{equation}
close to the  values reported by
Gonz{\'a}lez {\em et al.} (although their surmise that
$P/P_{\rm Poisson}$ is a simple fraction $1/d$ for $d=2$, $3$ is not
borne out by this exact calculation).
By including the effects of a finite temperature and screening length,
we obtain excellent agreement with the electric field profiles in the 
simulations (which could not be achieved in the drift approximation
of Ref.\ \cite{Beenakker1998}), and determine the conditions for
space-charge limited conduction.
We also go beyond previous work by calculating to what extent the
shot-noise suppression factor varies with the energy-dependence 
of the scattering rate.
(This breakdown of universality was anticipated
in Refs.\ \cite{Beenakker1998} and \cite{Nagaev1998}.)

The paper is organized as follows.
The kinetic theory is introduced in Sec.~\ref{sec:sec1},
where we summarize the basic equations and emphasize the differences
with the degenerate case.
In Sec.~\ref{sec:scl} we formulate the problem for the regime of
space-charge limited conduction.
In Sec.~\ref{sec:indep} we solve the kinetic equations for
the case of an energy-independent
scattering rate and compare with the Monte Carlo simulations
\cite{Gonzalez}.
We study separately the capacitance fluctuations.
The effect  of deviations from the
conditions of space-charge
limited conduction is also investigated.
Energy-dependence in the scattering rate
is considered
in Sec.~\ref{sec:dep}. 
We conclude in Sec.~\ref{sec:concl} with a discussion
of the experimental observability in  connection with
electron-phonon scattering.

\section{Kinetic theory}
\label{sec:sec1}
\subsection{Boltzmann-Langevin equation}
\label{sec:bl}

Our starting point is the same kinetic theory
\cite{Kadomtsev,Kogan,Gantsevich,Book}
used to study shot noise in degenerate conductors
\cite{reviews,Nagaev,many1,many2,many3,many3a,many4,many4a,many5}.
We summarize the basic equations, emphasizing the differences in the
non-degenerate case.
The density
$f({\bf r}, {\bf p}, t)$ of carriers
at position ${\bf r}$ and momentum ${\bf p}=m{\bf v}$ at time $t$
(where $m$ is the effective mass and {\bf v} the velocity)
satisfies the Boltzmann-Langevin equation
\begin{equation}
\label{eq:bl}
\left[\frac{\partial }{\partial t}
+{\bf v}\cdot\frac{\partial}{\partial \bf r}
+e{\bf E}({\bf r},t)\cdot\frac{\partial}{\partial \bf p}
\right]f({\bf r},{\bf p},t)={\cal S}+\delta J 
.
\end{equation}
Here ${\bf E}({\bf r},t)$ is the electric field
(we take the charge of
the carriers  positive),
${\cal S}({\bf r},{\bf p},t)$ is the collision integral,
and $\delta J({\bf r},{\bf p}, t)$ is a fluctuating source
(or ``Langevin current'').
The collision integral describes the average effect of
elastic impurity scattering,
\begin{eqnarray}
\lefteqn{{\cal S}({\bf r},p{\bf \hat n},t)}
\nonumber\\
&&\quad
=\int\frac{{\rm d}{\bf \hat n}'}{\Omega}
W_\varepsilon({\bf \hat n}\cdot{\bf \hat n}')
[f({\bf r},p{\bf {\hat n}}',t)-f({\bf r},p{\bf \hat n},t)] 
.
\label{eq:collint}
\end{eqnarray}
The integral over the direction ${\bf \hat
n}={\bf p}/p$ of the momentum
extends over the surface of the unit sphere
in $d$ dimensions, with surface
area $\Omega= 2\pi^{d/2}/\Gamma(\case 12 d)$.
The scattering rate
$W_\varepsilon({\bf \hat n}\cdot{\bf \hat n}')$
depends on the kinetic energy $\varepsilon=p^2/2m$ and
on the scattering angle 
${\bf \hat n}\cdot{\bf \hat n}'$. The effective mass $m$ is assumed to
be energy independent.

The stochastic Langevin current $\delta J$
vanishes on average, $\overline{\delta J} =0$, and has correlator 
\cite{Kogan}
\begin{eqnarray}
&&\overline{\delta J({\bf r},{\bf p},t)
\delta J({\bf r}',{\bf p}',t') }=
\delta({\bf r}-{\bf r}')
\delta(t-t')
\delta(\varepsilon-\varepsilon')\frac 1{\nu(\varepsilon)}
\nonumber
\\
&&
\quad
{}\times
\Big[
\delta({\bf \hat n}-{\bf \hat n}')
\int {\rm d}{\bf \hat n}''\,
W_\varepsilon({\bf \hat n}\cdot{\bf \hat n }'')
(\bar f+\bar f'' -2\bar f\bar f'')
\nonumber
\\
&&
\quad\qquad
{}-
W_\varepsilon({\bf \hat n}\cdot{\bf \hat n}')
(\bar f+\bar f'-2\bar f \bar f')
\Big] 
,
\label{eq:correl}
\end{eqnarray}
determined by the mean occupation number ${\bar f}$. (We abbreviated
$\bar f'=\bar f({\bf r},{\bf p}',t)$ and analogously for $\bar f''$.)
The density of states in $d$ dimensions is
$
\nu(\varepsilon)=
m\Omega (2m\varepsilon)^{d/2-1}$,
where we set Planck's constant $h\equiv 1$.

A non-degenerate electron gas is characterized by ${\bar f}\ll 1$.
In contrast to the degenerate case, the Pauli exclusion principle
is then of no effect. One consequence is that we may
omit the terms quadratic in ${\bar f}$ in the correlator (\ref{eq:correl}).
A second consequence is that deviations from equilibrium are no longer
restricted to a narrow energy range around the Fermi level, but extend
over a broad range of $\varepsilon$.  One can not, therefore, eliminate
$\varepsilon$ as an independent variable from the outset, as in
the degenerate case.

\subsection{Diffusion approximation}
\label{sec:dd}

We assume that the elastic mean free path is short compared to the 
dimensions  of the conductor, so that we can make the
diffusion approximation. This consists in
keeping only the two leading terms,
\begin{equation}
\label{eq:expansion}
f({\bf r},{\bf \hat n}\sqrt{2m\varepsilon},t)=
{\cal F}({\bf r},\varepsilon,t)+{\bf \hat n}\cdot
{\bf f}({\bf r},\varepsilon,t)
 ,
\end{equation}
of a multipole expansion in the momentum direction $\bf \hat n$.
We substitute Eq.~(\ref{eq:expansion})
into the Boltzmann-Langevin equation (\ref{eq:bl})
and integrate over ${\bf \hat n}$ to obtain the
continuity equation,
\begin{equation}
\label{eq:continuity}
\frac {\partial}{\partial t}\rho({\bf r},\varepsilon,t)
+
\frac {\partial}{\partial {\bf r}}\cdot{\bf j}({\bf r},\varepsilon,t)
+e{\bf E}({\bf r},t)\cdot
\frac {\partial}{\partial \varepsilon}{\bf j}({\bf r},\varepsilon,t)
=0
,
\end{equation}
for the energy-resolved  charge and current densities
\begin{eqnarray}
\label{eq:cc2}
\rho({\bf r},\varepsilon,t)&=&e \nu(\varepsilon)
{\cal F}({\bf r}, \varepsilon,t) 
,\\
{\bf j}({\bf r},\varepsilon,t)&=& \frac 1 d e v \nu(\varepsilon)
{\bf f}({\bf r}, \varepsilon,t)  
,
\end{eqnarray}
with $v=\sqrt{2\varepsilon/m}$.
In the zero-frequency limit we may omit the time derivative in
Eq.~(\ref{eq:continuity}).

Multiplication by ${\bf \hat n}$ followed by integration gives
a second relation between $\rho$ and ${\bf j}$ \cite{neglect},
\begin{eqnarray}
{\bf j}({\bf r},\varepsilon,t)
&=&
-D(\varepsilon)
\frac {\partial}{\partial {\bf r}}\rho({\bf r},\varepsilon,t)
\nonumber
\\
&&
{}-\sigma(\varepsilon){\bf E}({\bf r},t)
\frac {\partial}{\partial \varepsilon}
{\cal F}({\bf r},\varepsilon,t)
+
\delta{\bf J}({\bf r},\varepsilon,t)
 ,
\label{eq:dd}
\end{eqnarray}
a combination of Fick's law and Ohm's law with a fluctuating
current source.
The conductivity
$\sigma(\varepsilon)=e^2\nu(\varepsilon) D(\varepsilon)$
is the product
of the
density of states and the
diffusion constant
$D(\varepsilon)= v^2 \tau/d=(2\varepsilon/m d)\tau(\varepsilon) $.
The scattering rate is given by
\begin{equation}
\frac 1 {\tau(\varepsilon)} =
\int \frac{{\rm d}{\bf \hat n}'}{\Omega}W_\varepsilon
({\bf \hat n}\cdot{\bf \hat n}')
(1- {\bf \hat n}\cdot {\bf \hat n}') 
.
\end{equation}
The energy-resolved Langevin current
\begin{equation}
\delta{\bf J}({\bf r},\varepsilon,t)=
e\tau(\varepsilon) v\nu(\varepsilon)\int\frac{{\rm d}{\bf \hat n}}{\Omega}
{\bf \hat n}\,\delta J({\bf r},{\bf \hat n}\sqrt{2m\varepsilon},t)
\end{equation}
is correlated as
\begin{eqnarray}
\lefteqn{\overline{\delta J_l({\bf r},\varepsilon,t)
\delta  J_m({\bf r}',\varepsilon',t') }
}
\nonumber
\\
&&
{}=
2\sigma(\varepsilon)\bar {\cal F}({\bf r},\varepsilon,t)
\delta_{lm}\delta({\bf r}-{\bf r}')\delta(t-t')\delta(\varepsilon-
\varepsilon')
 ,
\label{eq:correl2}
\end{eqnarray}
where we have omitted terms quadratic in $\bar {\cal F}$.

These kinetic equations should be solved together with the
Poisson equation
\begin{equation}
\label{eq:poisson}
\kappa\frac{\partial }{\partial {\bf r}}\cdot{\bf E}({\bf r},t)
=\rho({\bf r},t)-\rho_{\rm eq} 
,
\end{equation}
with $\rho({\bf r},t)=\int{\rm d}\varepsilon \,
\rho({\bf r},\varepsilon,t)$ the integrated charge density,
$\kappa$ the dielectric constant, and $\rho_{\rm eq}$ the mean
charge density in equilibrium.
The Langevin current $\delta J$ induces fluctuations
in $\rho$ and hence in $\bf E$. The need to take the fluctuations in the
electric field into account selfconsistently is a severe complication of the
problem.

\subsection{Slab geometry}
\label{sec:geom}
\begin{figure}
\epsfxsize\columnwidth
\hspace*{.4cm}\epsffile{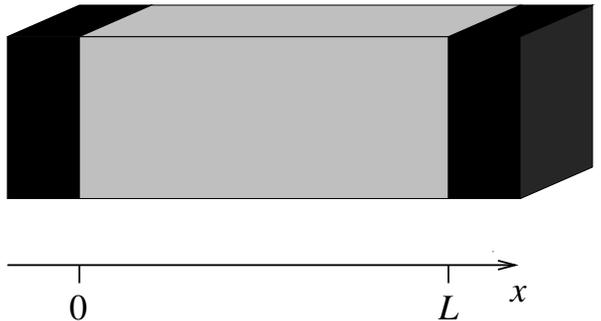}
\caption{
\narrowtext
Semiconducting slab (grey)
between two metal contacts (black)  at $x=0$ and $x=L$.
The $(d-1)$-dimensional 
cross-sectional area is $A$.
The current flows from left to right in response to a voltage $V$
applied between the contacts.
}
\label{fig:fig1}
\end{figure}

We consider the slab geometry of Fig.~\ref{fig:fig1},
consisting of a semiconductor aligned along the $x$-axis
with  uniform cross-sectional area $A$.
A non-fluctuating potential difference $V$ is maintained between
the metal contacts at $x=0$ and $x=L$, with the current source at
$x=0$. The contacts are in equilibrium at temperature $T$.
It is convenient
to integrate over energy and the coordinates ${\bf r}_\perp$ perpendicular to
the $x$-axis.
We define the linear charge density
$ \rho(x,t)=\int{\rm d}{\bf r}_\perp\, \rho({\bf r},t)$
and the currents
$I(t)=  \int{\rm d}{\bf r}_\perp\,\int{\rm d}\varepsilon\,
j_x({\bf r},\varepsilon,t)$  and
$\delta J(x,t)=  \int{\rm d}{\bf r}_\perp\,\int{\rm d}\varepsilon\,
\delta J_x({\bf r},\varepsilon,t)$.
The current $I$ is $x$-independent in the zero-frequency limit
because of the continuity equation
(\ref{eq:continuity}). 
We also define the electric field profile
$E(x,t)= A^{-1}\int{\rm d}{\bf r}_\perp\,E_x({\bf r},t)$.
The vector ${\bf r}_\perp$ of transverse coordinates has $d-1$ dimensions.
The physically relevant case is $d=3$, but in computer simulations one can consider other
values of $d$. For example, in Ref.~\cite{Gonzalez} the case $d=2$ was also studied,
corresponding to a hypothetical ``Flatland'' \cite{flatland}. To compare with the
simulations, we will also consider arbitrary $d$. 

For any $d$
the fluctuating Ohm-Fick law (\ref{eq:dd})
takes the one-dimensional form
\begin{eqnarray}
\label{eq:dd2new}
I(t)&=&-\frac{\partial}{\partial x}
\int{\rm d}{\bf r}_{\perp}\int {\rm d}\varepsilon \,
D(\varepsilon)\rho({\bf r},\varepsilon,t)
\nonumber
\\
&&
{}+E(x,t)
\int{\rm d}{\bf r}_{\perp}\int {\rm d}\varepsilon \,
{\cal F}({\bf r},\varepsilon,t)
\frac{\rm d}{{\rm d}\varepsilon}\sigma(\varepsilon)
\nonumber
\\
&&
{}
+\delta J(x,t) 
,
\end{eqnarray}
where we
used that the averages
of $\cal F$ and ${\bf E}$ depend on $x$ only and neglected terms quadratic
in the fluctuations.
The Poisson equation (\ref{eq:poisson}) becomes
\begin{equation}
\label{eq:poisson2}
\kappa A \frac{\partial}{\partial x}E(x,t)
=\rho(x,t)-A\rho_{\rm eq} 
,
\end{equation}
and the correlator (\ref{eq:correl2}) becomes
\begin{eqnarray}
\label{eq:correl3}
\lefteqn{\overline{\delta {J}(x,t)
\delta  {J}(x',t') }
}\nonumber\\
&&\quad =
2 A \delta(t-t')\delta(x-x')
\int{\rm d}\varepsilon\,
\sigma(\varepsilon)\bar {\cal F}(x,\varepsilon)
 .
\end{eqnarray}
Our problem is to compute from Eqs.~(\ref{eq:dd2new}) -- (\ref{eq:correl3})
the shot-noise power (\ref{eq:powerdef}).

\subsection{Energy-independent scattering time}
\label{sec:indepdd}
The Ohm-Fick law (\ref{eq:dd2new}) simplifies in the model of an
energy-independent scattering time $\tau(\varepsilon)\equiv \tau$.
Then the derivative of the conductivity ${\rm d}\sigma/{\rm
d}\varepsilon=e\mu\,\nu(\varepsilon)$ is proportional to the density of
states, and contains the energy-independent mobility $\mu=e\tau/m$.
Eq.~(\ref{eq:dd2new}) becomes 
\begin{eqnarray}
\label{eq:dd2}
I(t)&=&-\frac{\partial}{\partial x}
\int{\rm d}{\bf r}_{\perp}\int {\rm d}\varepsilon \,
D(\varepsilon)\rho({\bf r},\varepsilon,t)
\nonumber
\\
&&
{}+\mu\, \rho(x,t)E(x,t)+\delta
J(x,t) 
.
\end{eqnarray}
The drift term has now the same form $\mu\rho E$ as for inelastic
scattering \cite{ineldrift}.
This simple form does not hold for the more general case of
energy-dependent elastic scattering. 

\section{Space-charge limited conduction}
\label{sec:scl}
For a large voltage drop $V$ between the two metal contacts
and a high carrier density $\rho_c$ in the contacts,
the charge injected into the semiconductor is much higher than the
equilibrium charge $\rho_{\rm eq}$, which can then be neglected.
For sufficiently high $V$ and $\rho_c$ the system
enters the regime of space-charge limited conduction \cite{Lampert},
defined by the boundary condition 
\begin{equation}
\label{eq:bc1a}
E(x,t)=0\quad\mbox{at}\quad x=0 
.
\end{equation}
Eq.~(\ref{eq:bc1a}) states that the space charge 
$Q=\int_0^L\rho(x)\,{\rm d}x$ in the semiconductor is precisely
balanced by the surface charge at the current drain.
The accuracy of this boundary condition at finite $V$ and $\rho_c$
is examined in Sec.~\ref{sec:comp}.
At the drain we have the absorbing boundary condition
\begin{equation}
\label{eq:bc2a}
\rho(x,t)=0\quad\mbox{at} \quad x=L 
.
\end{equation}
With this boundary condition we again 
neglect $\rho_{\rm eq}$.

To determine the electric field inside the semiconductor we proceed as
follows. The potential gain $-e\phi(x,t)$ (with $E=-\partial
\phi/\partial x$) dominates over the initial thermal
excitation energy of order $kT$
(with Boltzmann's constant $k$)
almost throughout the whole
semiconductor; only close
to the current source
(in a thin boundary layer)
this is not the case.
We can therefore approximate the kinetic energy $\varepsilon\approx-e\phi$
and introduce 
this into $D(\varepsilon)$ and ${\rm d}\sigma/{\rm d}\varepsilon$.
We assume a power-law energy dependence of the scattering
time $\tau=\tau_0\varepsilon^\alpha$. Then
$D(\varepsilon)=(2\tau_0/md)\varepsilon^{\alpha+1}\approx
-(2\mu_0/d)(-e)^\alpha\phi^{\alpha+1}$ and 
${\rm d}\sigma/{\rm d}\varepsilon=(2\alpha+d)(\tau_0/md)e^2\varepsilon^\alpha
\nu(\varepsilon)\approx-(2\alpha+d)(\mu_0/d)(-e)^{\alpha+1}\phi^{\alpha}
\nu(\varepsilon)$, where we have defined $\mu_0=e\tau_0/m$.
Substituting
into Eq.~(\ref{eq:dd2new}) and using the Poisson equation
$-\kappa A {\partial^2}\phi/\partial x^2=\rho$ 
we find the third-order, non-linear, inhomogeneous
differential equation
\begin{eqnarray}
&&(2\alpha+d)\phi^\alpha\frac{\partial}{\partial x}
\left(
\frac{\partial \phi}{\partial x}
\right)^2
-4 
\frac{\partial}{\partial x}
\left(
\phi^{\alpha+1}
\frac{\partial^2 \phi}{\partial x^2}
\right)
\nonumber
\\
&&\qquad=\frac{2d}{(-e)^\alpha\mu_0 \kappa A}[I(t)-\delta{J}(x,t)]
\label{eq:dgl}
\end{eqnarray}
for the potential profile $\phi(x,t)$.

Since the potential difference $V$ between source and drain does not
fluctuate, we have the two boundary conditions
\begin{eqnarray}
\label{eq:bc3}
\phi(x,t)=&0~~&\quad\mbox{at}
\quad x=0 
,\\
\phi(x,t) =&-\! V &\quad\mbox{at}
\quad x=L 
.
\label{eq:bc4}
\end{eqnarray}
Eqs.~(\ref{eq:bc1a}) and (\ref{eq:bc2a}) imply
two additional boundary conditions,
\begin{eqnarray}
\label{eq:bc1}
\frac{\partial }{\partial x}\phi(x,t)=0&\quad\mbox{at}& \quad x=0 
,\\
\frac{\partial^2}{\partial x^2}\phi(x,t)=0&\quad\mbox{at}& \quad x=L 
.
\label{eq:bc2}
\end{eqnarray}

We will now solve this boundary value problem for $\phi=\bar \phi+\delta\phi$,
first for the mean and then for the fluctuations, in both cases
neglecting terms quadratic in $\delta\phi$.
The case $\alpha=0$ of an energy-independent scattering time is
considered first, in Sec.~\ref{sec:indep}. The more complicated case of
non-zero $\alpha$ is treated in Sec.~\ref{sec:dep}.

\section{Energy-independent scattering time}
\label{sec:indep}

\subsection{Average profiles}
\label{sec:mean}
For $\alpha=0$ the averaged equation (\ref{eq:dgl}) can be integrated once
to obtain the second-order differential equation
\begin{equation}
\label{eq:mean}
\left(\frac{{\rm d}\bar\phi}{{\rm d}x}\right)^2-\frac{4}{d}
\bar\phi\frac{{\rm d}^2\bar\phi}{{\rm d}x^2}=\frac{2\bar I}{\mu\kappa A}x
\end{equation}
for the mean potential $\bar \phi(x)$.
In this case of an energy-independent scattering time
$\tau(\varepsilon)\equiv \tau$ we may identify $\mu_0$ with the
mobility $\mu=e\tau/m$ introduced in Sec.~\ref{sec:indepdd}.
No integration constant appears in Eq.~(\ref{eq:mean}), since only then
the boundary conditions (\ref{eq:bc3}) and (\ref{eq:bc1}) at $x=0$
can be fulfilled simultaneously.
In Ref.~\cite{Beenakker1998} 
the second term on the left-hand-side of
Eq.~(\ref{eq:mean}) (the diffusion term)
was neglected relative to the first term (the drift term).
This approximation
is rigorously justified only in the formal limit $d\to\infty$.
It has the drawback of reducing the
order of the equation by one, so that no longer all boundary
conditions can be fulfilled. Although the solution
in Ref.~\cite{Beenakker1998} violates the absorbing boundary condition
(\ref{eq:bc2}), the final result for the shot-noise power turns out to
be close to the exact result obtained here.

Before 
solving this non-linear differential
equation exactly, we discuss two scaling
properties that help us along the way.
Note first that the current $\bar I$ can be scaled away by the substitution
\begin{equation}
\label{eq:rescpot}
\bar \phi(x)= -\left(\frac{2\bar I}{\mu \kappa A}\right)^{1/2} \chi(x)
. 
\end{equation}
Secondly, 
each solution
$\chi(x)$ of 
\begin{equation}
\label{eq:mean2}
\left(\frac{{\rm d}\chi}{{\rm d}x}\right)^2-\frac{4}{d}
\chi\frac{{\rm d}^2\chi}{{\rm d}x^2}=x
\end{equation}
[the rescaled Eq.~(\ref{eq:mean})]
generates a one-parameter family of solutions
$\lambda^{3/2} \chi(x/\lambda)$.
Thus, if we find a solution
which fulfills the three boundary
conditions $\chi(0)=0$, $\chi'(0)=0$,
$\chi''(1)=0$ (primes denoting differentiation with respect to $x$),
then the potential
\begin{equation}
\bar\phi(x)=-\left(\frac{2\bar I L^3}{\mu\kappa A}\right)^{1/2} \chi(x/L)
\end{equation}
solves
Eq.~(\ref{eq:mean}) with boundary conditions (\ref{eq:bc3}),
(\ref{eq:bc1}), and (\ref{eq:bc2}). The remaining boundary condition
(\ref{eq:bc4}) determines the current-voltage characteristic
\begin{equation}
\label{eq:iv}
\bar I(V)=\frac {\mu\kappa A}{2L^3} \left(\frac{V}{\chi(1)}\right)^2 
.
\end{equation}
The quadratic dependence of $\bar I$ on $V$ is the
Mott-Gurney law of space-charge limited conduction
\cite{Mott}.

We now construct a solution $\chi(x)$.
One obvious solution is
$\chi_0(x)= a_0 x^{3/2}$, with 
\begin{equation}
\label{eq:theta0}
a_0=\frac 23 \left(1 - \frac{4}{3d} \right)^{-1/2}
 .
\end{equation}
This solution satisfies the boundary conditions at $x=0$, but 
$\chi''_0(x)\neq 0$ for any finite $x$.
Close to the singular point $x=0$ any solution
will approach $\chi_0(x)$ provided that $d> 4/3$.
Let us discuss first this range of
$d$, containing the physically relevant dimension $d=3$.

We substitute into Eq.~(\ref{eq:mean2}) the ansatz 
\begin{equation}
\chi(x)=\sum_{l=0}^\infty
a_l x^{\beta l+3/2}
 ,
 \label{eq:chiexp}
\end{equation}
consisting of $\chi_0(x)$ times a power series in $x^\beta$, with $\beta$ a
positive power to be determined.
This ansatz proves fruitful since
both terms on the left-hand-side of Eq.~(\ref{eq:mean2}) give the same powers
of $x$, starting with order $x^1$ in coincidence with the right-hand-side.
Power matching gives Eq.~(\ref{eq:theta0}) for
$a_0$ and for $l\geq 1$ the conditions
\begin{equation}
\sum_{m=0}^l b_{lm} a_{m}a_{l-m}=0 
,
\end{equation}
\begin{equation}
b_{lm}=\frac 9 4-\frac 3d +
\left(3-\frac{8}d\right) m\beta-\left(1+\frac4d\right)m^2\beta^2
+ml\beta^2
 .
\end{equation}
The relation with $l=1$ is special: It determines the power $\beta$,
\begin{equation}
\label{eq:p}
\beta^2+\beta\left(2-\frac 34 d\right)+\frac 32-\frac 98 d=0
 ,
\end{equation}
but leaves the coefficient $a_1$ as a free parameter [to be determined
by demanding that $\chi''(1)=0$].
The positive solution of Eq.~(\ref{eq:p}) is
\begin{equation}
\beta=\frac 3 8 d-1+\frac  18 \sqrt{9d^2+24 d-32} 
.
\end{equation}
We find  $\beta=(\sqrt {13}-1)/4$ for $d=2$ and  $\beta=3/2$ for $d=3$.
For $l\ge 2$ we solve for $a_l$ to obtain the recursion relation
\begin{equation}
a_l=-\frac{
\sum_{m=1}^{l-1}
b_{lm}a_ma_{l-m}
}{
(b_{ll}+b_{l0})a_0
}
 .
\end{equation}

\begin{figure}
\epsfxsize\columnwidth
\epsffile{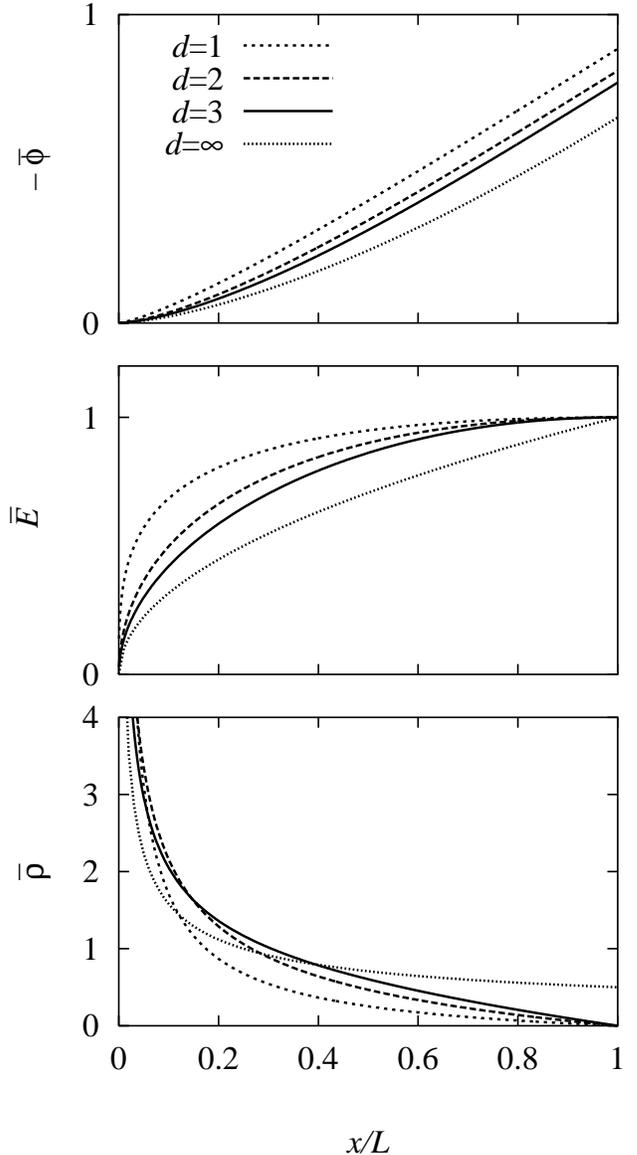}
\caption{
\narrowtext
Profile of the mean electrical potential $\bar \phi$ [in units of
$(2\bar I L^3/\mu\kappa A)^{1/2}$], 
the electric field $\bar E$ [in units of $(2\bar I L/\mu\kappa A)^{1/2}$],
and the charge density $\bar \rho$
[in units of $(2\bar I \kappa/\mu L A)^{1/2}$], following from
Eq.~(\protect\ref{eq:mean}) for different values of $d$.
The drift approximation of
Ref.~\protect\cite{Beenakker1998} corresponds to
the case $d=\infty$ in this plot.
}
\label{fig:fig2}
\end{figure}

\noindent
Interestingly enough, the power series terminates for $d=12/5$,
and the solution for this dimension 
is $\chi(x)=x^{3/2}-\case 15 x^{5/2}$.
For arbitrary dimension $d>4/3$,
the coefficients $a_l$ fall off with $l$, the more rapidly so
the larger $d$.
We find numerically that the solution with
$\chi''(1)=0$ has
$a_1=0.3261$ for $d=2$ and
$a_1=0.1166$ for $d=3$.

For $d<4/3$ 
we substitute into Eq.~(\ref{eq:mean2}) the ansatz
\begin{equation}
\chi(x)=\sum_{l=0}^\infty
c_l x^{\gamma l+(3-\gamma)/2}
 ,
\end{equation}
with $\gamma=(4-3d)/(4-d)$.
Now the coefficient $c_0$ is free. Power matching gives further
\begin{equation}
c_1=-\frac{d}{4\gamma(\gamma+1)}
 ,
\end{equation}
and the recursion relation
\begin{equation}
c_l=-\frac{
\sum_{m=1}^{l-1}
d_{lm}c_mc_{l-m}
}{
(d_{ll}+d_{l0})c_0
}
 ,
\end{equation}
\begin{equation}
d_{lm} = \left(\gamma m+\frac{3-\gamma}{2}\right)
\left[\gamma (l-m)-\frac 4d \gamma m\right]
 ,
\end{equation}
for coefficients with $l\ge 2$.
For $d=1$ the solution with
$\chi''(1)=0$ has $c_0=1.3628$.

In Fig.~\ref{fig:fig2}
the profiles
of the potential $\bar\phi\propto \chi$, the electric field $\bar E\propto
\chi'$, and the charge density
$\bar\rho\propto\chi''$ are plotted for $d=1$, $2$, $3$.
We also show the result for $d=\infty$, corresponding to the
drift approximation of Ref.~\cite{Beenakker1998}.
The coefficient 
$\chi(1)$ appearing in the current-voltage characteristic
(\ref{eq:iv}) can be read off from this plot.
We find $\chi(1)=8/9$ for $d=1$,
$\chi(1)=0.8180$ for $d=2$, and $\chi(1)=0.7796$ for $d=3$.
The limiting value for $d=\infty$ is $\chi(1)=2/3$.

\subsection{Fluctuations}
\label{sec:fluc}

For the fluctuations it is again convenient to work with
the rescaled mean potential (\ref{eq:rescpot}).
We rescale the fluctuations in the same way,
\begin{equation}
\delta \phi(x,t)=-\left(\frac{2\bar I}{\mu\kappa A}\right)^{-1/2}\psi(x,t)
 .
\end{equation}
We linearize
Eq.~(\ref{eq:dgl}) with $\alpha=0$ around the mean values and
integrate once to obtain
the second-order inhomogeneous linear differential equation
\begin{eqnarray}
{\cal L}[\psi]&=&
-\left(\frac 4 d \chi\right)
\frac{{\partial}^2\,\psi}{{\partial}x^2}
+
\left(2\frac{{\rm d}\chi}{{\rm d}x}\right)
\frac{{\partial}\,\psi}{{\partial}x}
-\left(\frac 4 d
\frac{{\rm d}^2\chi}{{\rm d}x^2}\right)\psi
\nonumber
\\
&=&
\int_0^x{\rm d}x'\,\frac{\delta I(t)-\delta {J}(x',t)}{\bar I}
\label{eq:fluct}
 .
\end{eqnarray}
The integration constant vanishes as a
consequence of the boundary condition
\begin{equation}
\label{eq:bc5}
\psi(x,t)=0\quad\mbox{at} \quad x=0
\end{equation}
and the requirement that the fluctuating electric field
${\partial \psi}/{\partial x}$ stays finite at $x=0$.
[The latter condition actually implies $\partial\psi/{\partial x}=0$ at $x=0$.]
We will solve  Eq.~(\ref{eq:fluct}) with the additional condition
of a non-fluctuating voltage,
\begin{equation}
\psi(x,t) =0 \quad\mbox{at} \quad x=L.
\label{eq:bc8}
\end{equation}
The remaining constraint
\begin{equation}
\label{eq:bc6}
\frac{\partial^2}{\partial x^2}\psi(x,t)=0\quad\mbox{at} \quad x=L
\end{equation}
(the absorbing boundary condition)
will be used later to relate $\delta I$ to  $\delta {J}$.

We need the Green function $G(x,x')$,
satisfying for each $x'$ the equation ${\cal
L}[G(x,x')]=\delta(x-x')$.
In view of Eq.~(\ref{eq:mean2})
for the mean potential one recognizes
\begin{equation}
\psi_1(x)= 3\chi(x)-
2x\,\frac{{\rm d}}{{\rm d}x}\chi(x) 
\end{equation}
as a solution ${\cal L}[\psi_1]=0$ which
already satisfies Eq.~(\ref{eq:bc5}).
Using a standard prescription \cite{Ince} we find from $\psi_1(x)$ a second,
independent, homogeneous solution
\begin{equation}
\psi_2(x)= \psi_1(x)\int_x^L\!\!{\rm d}x'\,
\frac{\chi^{d/2}(x')}
{\psi_1^2(x')}
 ,
\end{equation}
which fulfills Eq.~(\ref{eq:bc8}).
The Wronskian is 
\begin{equation}
\psi_1(x)\frac{\rm d}{{\rm d}x}
\psi_2(x)-\psi_2(x)\frac{\rm d}{{\rm d}x}\psi_1(x)=
-\chi^{d/2}(x) 
.
\end{equation}
The Green function contains also the factor $-4\chi/d$ that
appears in Eq.~(\ref{eq:fluct}) in front of the second-order
derivative of $\psi$.
We find
\begin{eqnarray}
G(x,x')&=&
\frac{d}{4\chi^{d/2+1}(x')}\left[
\Theta(x-x') \psi_2(x)\psi_1(x')
\right.
\nonumber 
\\
&&\quad \left.
{}+
\Theta(x'-x) \psi_1(x)\psi_2(x')
\right] 
,
\end{eqnarray}
where $\Theta(x)=1$ for $x>0$ and $\Theta(x)=0$ for $x<0$.

The solution of the inhomogeneous equation
(\ref{eq:fluct}) with boundary
conditions (\ref{eq:bc5}), (\ref{eq:bc8}) is then
\begin{equation}
\psi(x,t)=
\int_0^L{\rm d}x'\,G(x,x')
\int_0^{x'}{\rm d}x''\,
\frac {\delta I(t)-\delta {J}(x'',t)}{\bar I}
 .
\label{eq:psii}
\end{equation}
From the extra condition (\ref{eq:bc6}) we find
\begin{equation}
\delta I(t)=
{\cal C}^{-1}
\int_0^L{\rm d}x\,
\delta {J}(x,t){\cal G}(x)
 ,
\label{eq:deltai}
\end{equation}
with the definitions
\begin{equation}
{\cal C}= \left(\frac 32 \chi(L) -L^{3/2}\right)+
\frac d4\frac {\chi^{d/2}(L)}{\sqrt L}
\int_0^L\!\!{\rm d}x\,\frac{x \psi_1(x)}{\chi^{d/2+1}(x)}
 ,
\label{eq:calc}
\end{equation}
\begin{equation}
{\cal G}(x)=
\left(\frac{3\chi(L)}{2L}-\sqrt L\right)+
\frac d4\frac{\chi^{d/2}(L)}{\sqrt L}
\int_x^L\!\!{\rm d}x'\,
\frac{\psi_1(x')}{\chi^{d/2+1}(x')}
 .
\label{eq:calg}
\end{equation}
Eq.~(\ref{eq:deltai}) is the relation between the fluctuating total
current $\delta
I$ and the Langevin current $\delta J$ that we need
to compute the  shot-noise power.

\subsection{Shot-noise power}
\label{sec:power}

The shot-noise power is found by substituting Eq.~(\ref{eq:deltai})
into Eq.~(\ref{eq:powerdef}) and invoking the correlator
(\ref{eq:correl3}) for the Langevin current.
This results in 
\begin{eqnarray}
\label{eq:pres1}
&&P=2\int_0^L{\rm d}x\,\left(\frac{{\cal G}(x)}{\cal C}\right)^2
{\cal H}(x) 
,
\\
&&{\cal H}(x)=
2A\int{\rm d}\varepsilon\,\sigma(\varepsilon)
\bar{\cal F}(x,\varepsilon)
 .
\label{eq:calh}
\end{eqnarray}
In order to determine the
mean occupation number
$\bar{\cal F}(x,\varepsilon)$ out of equilibrium, 
it is convenient to change variables from
kinetic energy $\varepsilon$ to total energy $u=\varepsilon+e\bar\phi(x,t)$.
In the new variables $x$ and $u$
we find
from the 
kinetic equations (\ref{eq:continuity}) and (\ref{eq:dd})
\begin{eqnarray}
&&\frac {\partial}{\partial x}\bar j(x,u)=0 
,
\\
&&\bar j(x,u)=-\frac 1e\sigma[u-e\bar\phi(x)]
\frac{\partial }{\partial x}\bar{\cal
F}(x,u) 
.
\end{eqnarray}
The derivatives with respect to $x$ are taken at constant $u$.
The solution is
\begin{equation}
\label{eq:barf}
\bar{\cal F}(x,u)=
e\bar j(u)\int_x^L\frac{{\rm d}x'}{\sigma[u-e\bar\phi(x')]}
 ,
\end{equation}
where we used the absorbing boundary condition (\ref{eq:bc2})
(which implies
$\bar {\cal F}(L,u)=0$).

As before [in the derivation of Eq.~(\ref{eq:dgl}) from
Eq.~(\ref{eq:dd2new})]
we approximate $u-e\bar\phi(x)\approx 
-e\bar\phi(x)$ in the argument of $\sigma$.
(This is justified because $0<u\lesssim kT\ll eV$.)
Then $\bar {\cal F}(x,u)$ factorizes into a function of $x$ times a
function of $u$, and Eq.~(\ref{eq:calh}) gives
\begin{equation}
{\cal H}(x)=
2e\bar I\,\chi^{d/2}(x')
\int_x^L{\rm d}x'\,\chi^{-d/2}(x') 
,
\label{eq:calh2}
\end{equation}
where we expressed the result in terms of the rescaled potential $\chi$.
In this equation we recognize the  Poissonian shot-noise power $P_{\rm
Poisson}=2e\bar I$.

The integrals in the expressions (\ref{eq:calc}),
(\ref{eq:calg}), and
(\ref{eq:calh2})
for $\cal C$, $\cal G$,
and ${\cal H}$ can be performed 
with the help of the fact that $\chi$
solves the differential equation (\ref{eq:mean2}).
In view of this equation, 
\begin{eqnarray}
\label{eq:diff1}
\chi^{-d/2}&=&-\frac 4d
\frac{{\rm d}}{{\rm d}x}
\left(
\chi^{1-d/2}\chi''
\right)
 ,\\
\frac{x\chi'}{\chi^{d/2+1}}
&=&
-\frac 2 d \frac{\rm d}{{\rm d}x}
\left(
\chi^{-d/2}
{\chi'}^2
\right)
 ,
\\
\frac{x\psi_1}{\chi^{d/2+1}}
&=&\frac 4 d\frac{\rm d}{{\rm d}x}\left[\left(
x{\chi'}^2
-
\chi\chi'-x\chi\chi''
\right)\chi^{-d/2}
\right]
 ,
\end{eqnarray}
resulting in
\begin{eqnarray}
&&{\cal C}=\case 12 \chi(L) 
,
\\
&&{\cal H}(x)
=P_{\rm Poisson}\frac{4}{d}
\chi(x)\chi''(x)
,
\label{eq:hresult}
\\
&&{\cal G}(x)=
\frac 1{\sqrt L}\left[3 \chi(x)
\chi''(x)
-{\chi'}^2(x)\right]
\left(\frac{\chi(L)}{\chi(x)}\right)^{\case d2}
+\frac{3\chi(L)}{2L} 
.
\nonumber
\\
\end{eqnarray}

\begin{figure}
\epsfxsize\columnwidth
\epsffile{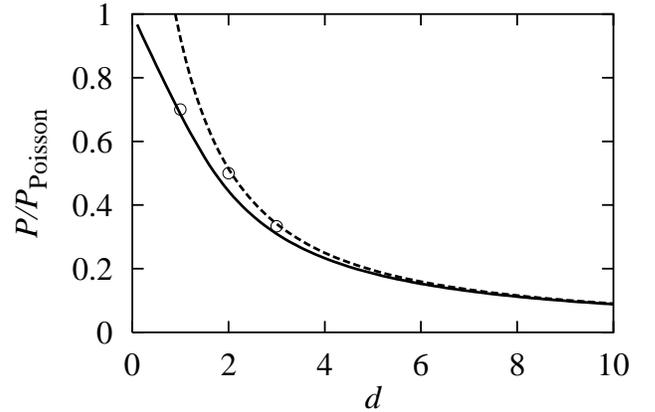}
\caption{
\narrowtext
Shot-noise power $P$ for an energy-independent scattering rate
as a function of $d$.
The exact result (full curve) is compared with the approximate result
(\protect\ref{eq:analyt}) (dashed curve).
Both curves approach $4/5d$ for $d\to\infty$.
The data points are the results of
numerical simulations \protect\cite{Gonzalez}.
}
\label{fig:fig3}
\end{figure}

Our final expression for the shot-noise power is
\begin{equation}
P= P_{\rm Poisson}\frac {32} {d}\frac 1{\chi^2(L)}
\int_0^L{\rm d}x\,{\cal G}^2(x)\chi(x)
\frac {{\rm d}^2}{{\rm d} x^2}\chi(x)
 .
\label{eq:powerres}
\end{equation}
The scaling properties of $\chi$ imply
that this result does not depend on the length $L$.
For $d=1$, 2, 3 it evaluates to
\begin{equation}
P/P_{\rm Poisson}=\left\{
\begin{array}{cc}
0.6857& \quad\mbox{for}\quad d=1,\\
0.4440& \quad\mbox{for}\quad d=2,\\
0.3097& \quad\mbox{for}\quad d=3.
\end{array}
\right. 
\end{equation}
In Fig.~\ref{fig:fig3} we plot Eq.~(\ref{eq:powerres}) as a function
of the dimension $d$ and compare it with the
approximate formula
(\ref{eq:analyt}),
obtained in Ref.~\cite{Beenakker1998} by neglecting
the diffusion term in Eq.~(\ref{eq:mean}).
The exact result (\ref{eq:powerres})
is smaller than the approximate result
(\ref{eq:analyt}) by about $10\%$, $15\%$, and $25\%$ for
$d=3$, $2$, and $1$, respectively. For $d\to\infty$, the drift approximation
that leads to Eq.~(\ref{eq:analyt}) becomes strictly justified, and $P/P_{\rm
Poisson}$ approaches $4/5d$.
The data points in Fig.~\ref{fig:fig3} are the result of the numerical
simulation \cite{Gonzalez}. The agreement with the theory presented
here is quite satisfactory, although our findings do not support
the conclusion of Ref.~\cite{Gonzalez} that $P=\case 13 P_{\rm Poisson}$
in three dimensions.

\subsection{Capacitance fluctuations}
\label{sec:vn}

The fluctuations $\delta I(t)$ 
in the current $I(t)$ are due in part to fluctuations
in the total charge 
$Q(t)=\int{\rm d}x\,\rho(x,t)$ in the
semiconductor. The contribution from this source to the current
fluctuations
is $\delta I_Q=(\delta Q/\bar Q)\bar I$. Fluctuations in the carrier
velocities account for the remaining current fluctuations $\delta I_V=\delta
I-\delta I_Q$.
Since the fluctuations in $Q$ could be measured capacitatively,
it is of interest to compute their magnitude separately.
Because we have assumed that there is no charge present in equilibrium
in the semiconductor, $Q(t)=C(t)V$ is directly proportional to the
applied voltage $V$. The proportionality constant $C(t)$ is the
fluctuating capacitance of the semiconductor.
(The voltage does not fluctuate.)

 With the Poisson equation
(\ref{eq:poisson2}) and the boundary condition
(\ref{eq:bc1a}) we have 
\begin{equation}
C(t)=\frac{\kappa A}{V} E(L,t).
\end{equation}
The correlator of the capacitance fluctuations,
\begin{equation}
\label{eq:cappower}
P_C=2\int_{-\infty}^\infty{\rm d}t\,
\overline{\delta C(0) \delta C(t)},
\end{equation}
is related to the correlator of $\delta I_Q$,
\begin{equation}
\label{eq:qpower}
P_Q=2\int_{-\infty}^\infty{\rm d}t\,
\overline{\delta I_Q(0) \delta I_Q(t)},
\end{equation}
by
$P_Q=(\mu \bar I V^2/2\kappa A L) P_C$.
We also define the correlators
\begin{eqnarray}
\label{eq:vpower}
P_V&=&2\int_{-\infty}^\infty{\rm d}t\,
\overline{\delta I_V(0) \delta I_V(t)},
\\
\label{eq:qvpower}
P_{QV}&=&4\int_{-\infty}^\infty{\rm d}t\,
\overline{\delta I_Q(0) \delta I_V(t)},
\end{eqnarray}
such that $P=P_Q+P_V+P_{QV}$.

In view of Eqs.~(\ref{eq:dgl}), (\ref{eq:fluct}) and the boundary
conditions (\ref{eq:bc4}), (\ref{eq:bc2}), one obtains $\bar E(L)$ and
$\delta E(L,t)$ as a function of $\delta I$ and $\delta J$,
and hence
\begin{eqnarray}
&&\delta I_N(t)=\frac 12 \left(\delta I(t)-\int_0^L \frac{{\rm d}x}{L}\,
\delta J(x,t)\right)
 ,
\\
&&\delta I_V(t)=\frac 12 \left(\delta I(t)+\int_0^L \frac{{\rm d}x}{L}\,
\delta J(x,t)\right)
 .
\end{eqnarray}
With the help of Eq.~(\ref{eq:deltai}) we find
\begin{eqnarray}
P_{C}&=&\case 14 (P+P_{J}-2P_{IJ}) 
,\\
P_{V}&=&\case 14 (P+P_{J}+2P_{IJ}) 
,\\
P_{CV}&=&\case 12 (P-P_{J}) 
,
\\
P_{IJ}&=&\frac {16}{d\chi(L)}
P_{\rm Poisson}\int_0^L\frac{{\rm d}x}{L}\,
{\cal G}(x)\chi(x) \frac {{\rm d}^2}{{\rm d} x^2}\chi(x) 
,
\\
P_{J} &=&\frac {8}d
P_{\rm Poisson}\int_0^L\frac{{\rm d}x}{L^2}\,
\chi(x)\frac {{\rm d}^2}{{\rm d} x^2}\chi(x) 
.
\end{eqnarray}

\begin{figure}
\epsfxsize\columnwidth
\epsffile{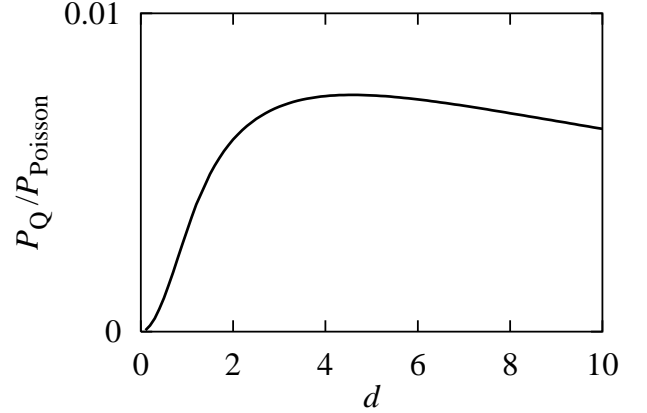}
\caption{
\narrowtext
Contribution $P_Q$ from charge fluctuations to the shot-noise power $P$.
The correlator $P_C$ of the capacitance fluctuations is related to $P_Q$ by
$P_C=(4e\kappa A L/\mu V^2) P_Q/P_{\rm Poisson}$.
}
\label{fig:fig4}
\end{figure}

\noindent
The integrals can be evaluated by using that $\chi(x)$ solves
Eq.~(\ref{eq:mean2}), with the result
\begin{eqnarray}
P_{IJ}&=&
4\frac{
48L^{-3/2}\chi(L)-d-36
}{(d+4)(1-5d)}P_{\rm Poisson} 
,
\\
P_{J} &=&
4\frac{2L^{-3/2}\chi(L)-1}{d+4}
P_{\rm Poisson}
 .
\end{eqnarray}

In Fig.~\ref{fig:fig4} the
correlator of the capacitance fluctuations is plotted 
as a function of $d$. For $d=3$ we find $P_C=0.0284 e\kappa AL/\mu V^2$.
The corresponding contribution $P_Q=0.0071P_{\rm Poisson}$ is relatively
small, being less than $3\%$ of the contribution from 
the velocity fluctuations $P_V=0.3076 P_{\rm Poisson}$.
(Incidentally, we find that charge and velocity fluctuations 
are anticorrelated, $P_{QV}=-0.0049 P_{\rm Poisson}$.) Our calculation
thus confirms the numerical finding of Ref.~\cite{Gonzalez}, that the
charge fluctuations are strongly suppressed as a result of Coulomb
repulsion. However, we do not find the exact cancellation of $P_Q$ and
$P_{QV}$ surmised in that paper.

\subsection{Effects of a finite voltage and carrier density}
\label{sec:comp}

For comparison with realistic systems
and with computer simulations
one has to account for a finite voltage $V$ and a finite carrier density
$\rho_c$ in the metal contacts.
The relevant parameters are 
the ratios
$L_c/L$ and $L_s/L$, with 
$L_c=(\kappa k T/e\rho_c)^{1/2}$ the Debye screening length in the contact
and $L_s=(\kappa V/\rho_c)^{1/2}$ the screening length in the semiconductor.
The theory of space-charge limited conduction applies to the regime $L\gg L_s\gg L_c$
(or $kT\ll eV$ and $\rho_c\gg \kappa V/L^2$ --- the combination $\kappa V/L^2$ 
characterizing the mean charge density in the semiconductor).
In this section we will show that, within this regime,
the effects of a finite voltage and carrier density are restricted to a
narrow boundary layer near the current source. We will examine the deviations
from the boundary condition (\ref{eq:bc1a}) and
compare with the numerical simulations \cite{Gonzalez}.

To investigate the accuracy of the boundary 
condition (\ref{eq:bc1a}) we start from the more fundamental 
condition of thermal equilibrium, 
\begin{equation}
\bar\rho(x,\varepsilon)=\frac{A\rho_c\nu(\varepsilon)\exp(-\varepsilon/kT)}
{\int_0^\infty {\rm d}\varepsilon'\,\nu(\varepsilon')\exp(-\varepsilon'/kT)}
\quad\mbox{at}\quad x=0.
\label{eq:thermeq}
\end{equation}
We keep the
absorbing boundary condition $\bar\rho(L,\varepsilon)=0$
at the current drain, since thermally excited carriers injected
from the contact at $x=L$ make only a small contribution to the
current when $eV\gg kT$.
To simplify the problem we assume that all carriers at the current
source have the same kinetic energy $\frac 12 d kT$, in essence replacing the
Boltzmann factor $\exp(-\varepsilon/kT)$ in Eq.~(\ref{eq:thermeq})
by a delta function at $\varepsilon=\frac d2 kT$. We restrict ourselves to the
physically relevant case $d=3$ and
substitute $\varepsilon=\frac 32 kT-e\bar\phi(x)$ in the argument
of $D(\varepsilon)$ in Eq.~(\ref{eq:dd2}).
Repeating the steps that resulted in Eq.~(\ref{eq:mean}),
we arrive at the differential equation 
\begin{equation}
\label{eq:finitet}
\left(\frac{{\rm d}\bar\phi}{{\rm d}x}\right)^2-\left(\frac{4}{3}
\bar\phi-2\frac{kT}{e}\right)
\frac{{\rm d}^2\bar\phi}{{\rm d}x^2}=\frac{2\bar I}{\mu\kappa A}
(x-\xi).
\end{equation}
In comparison to Eq.~(\ref{eq:mean}), an integration constant $\xi$
appears now
on the right-hand-side. This constant and the current $\bar I$ have to be 
determined from the four
boundary conditions $\bar \phi(0)=0$,
$\kappa\bar \phi''(0)=-\rho_c$, $\bar \phi(L)=-V$,  $\bar\phi''(L)=0$.

\begin{figure}
\epsfxsize\columnwidth
\epsffile{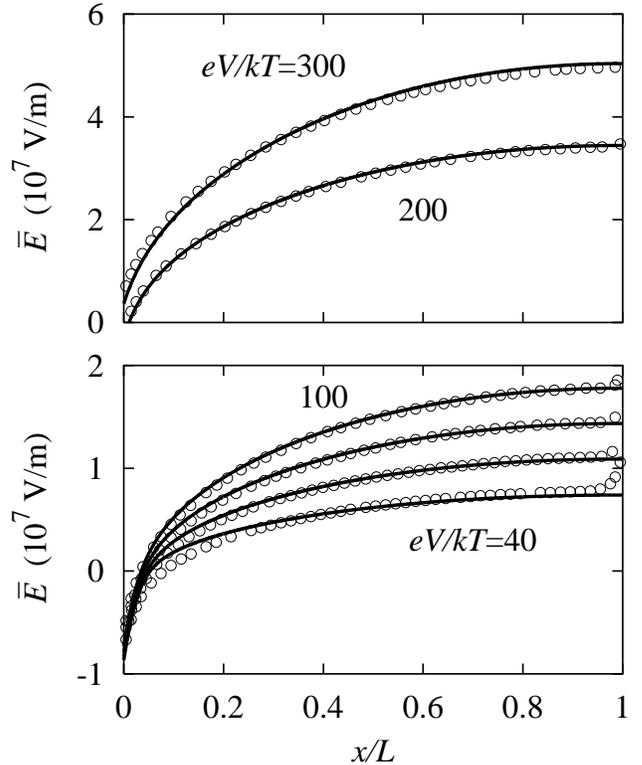}
\caption{
\narrowtext
Electric field profiles for $eV/kT=40$, 60, 80, 100, 200, 300,
at parameter values
$d=3$,
$T=300\,\mbox{K}$,
$\rho_c/e=10^{24}\,\mbox{m}^{-3}$,
$L=200\,\mbox{nm}$,
$\kappa=11.7\kappa_0$ (with $\kappa_0$ the dielectric constant of
vacuum).
The full curves follow from Eq.~(\protect\ref{eq:finitet}).
The data points are the result of numerical
simulations \protect\cite{gonzalezpriv}.
There are no fitting parameters in this comparison.
}
\label{fig:fig5}
\end{figure}

We have integrated
Eq.~(\ref{eq:finitet}) numerically. 
In Fig.~\ref{fig:fig5} we show the electric field for
$d=3$ and parameters as in the simulations of Ref.~\cite{Gonzalez},
corresponding to $L/L_c=48.9$ and $(L_s/L_c)^2=eV/kT$ ranging
between $40$ and $300$.
We find excellent agreement, 
the better so the larger $eV/kT$, {\em without any adjustable
parameter}.

The boundary  condition (\ref{eq:bc1a}) of zero electric field
at the current source assumes that the 
surface charge in the current drain is fully screened by the 
space charge in the semiconductor. With increasing
$eV/kT$ for fixed $L/L_c$ one observes in Fig.\ \ref{fig:fig5}
a transition from overscreening
($\bar E=0$ at a point inside the semiconductor) to underscreening
($\bar E$ extrapolates to zero at a point inside the metal
contact). We can approximate $\bar E(x)=-\bar\phi_0'(x-\xi)$,
where $\bar\phi_0$ solves Eq.~(\ref{eq:mean})
with the boundary conditions of space-charge limited conduction.
This is an excellent approximation for $eV/kT=200$ ($\xi/L=0.02$)
and $eV/kT=300$ ($\xi/L=-0.004$), practically indistinguishable
from the curves in 
Fig.\ \ref{fig:fig5} (top panel). 

To demonstrate analytically that space-charge limited
conduction is characterized by the conditions $L\gg L_s\gg L_c$,
we will now compute the width of the boundary layer
and show that it becomes $\ll L$ in this regime.
We need to distinguish between two length scales $\xi$ and $\xi'$
to fully characterize the boundary layer. The length $\xi$
determines the 
shift in the asymptotic solution
\begin{equation}
\bar\phi_{\rm asym}(x)= \bar\phi_0(x-\xi)+3kT/2e,
\label{eq:shift}
\end{equation}
while the size $\xi'$ characterizes the
range $0\le x \lesssim \xi'$ 
where the exact solution $\bar\phi(x)$ deviates substantially
from $\bar\phi_{\rm asym}(x)$.

The values of $\xi$ and $\xi'$ are 
found by comparing Eq.\ (\ref{eq:shift}) with the Taylor
series
\begin{equation}
\bar\phi(x)=-E_0x-
\frac{\rho_c}{\kappa} \frac{x^2}{2}+\phi_3 \frac{x^3}{6}
+{\cal O}(x^4).
\label{eq:blexp}
\end{equation}
The coefficients in the Taylor series are determined from 
Eq.~(\ref{eq:finitet}),
\begin{eqnarray}
&&E_0^2-2\frac{kT}{e}\frac{\rho_c}{\kappa}=-\xi\frac{2\bar I}{\mu\kappa A},
\label{eq:bl1}
\\
&&\frac 23 E_0\frac{\rho_c}{\kappa}+2\frac{kT}{e}\phi_3=
\frac{2\bar I}{\mu\kappa A}
,
\label{eq:bl2}
\end{eqnarray}
where $2\bar I/\mu\kappa A\approx V^2/L^3$ up to a coefficient of order unity
[cf.\ Eq.\ (\ref{eq:iv})].

We match the two functions (\ref{eq:shift}) and (\ref{eq:blexp}) at $x=\xi'$, 
demanding that potential and electric field are continuous at $x=\xi'$. These
two conditions determine  $\xi$ and $\xi'$.
Within the regime $L\gg L_s\gg L_c$ we
find two subregimes, depending on the relative magnitude
of $L_c/L$ and $(L_s/L)^4$. 
{\em Overscreening} occurs when $L_c/L\gg (L_s/L)^4$.
Then $E_0\approx -(2kT\rho_c/e\kappa)^{1/2}$,
$\phi_3\approx(2e/9kT)^{1/2}(\rho_c/\kappa)^{3/2}$,
and $\xi\approx\xi'={\cal O}(L_c)$.
The difference
$\xi'-\xi={\cal O}(L_s^4/L^3)\ll\xi$.
At the matching point,
$\bar\phi={\cal O}(kT/e)$,  $\bar E={\cal O}(V^2\kappa/\rho_cL^3)$, and
$\bar\rho={\cal O}(\rho_c)$.
{\em Underscreening} occurs when
$L_c/L\ll (L_s/L)^4$.
Then $E_0={\cal O} (V^2\kappa /\rho L^3)\ll V/L$,
$\phi_3={\cal O}(\rho_c^3L^3/\kappa^3V^2)$,
$\xi=-{\cal O}(L_s^4/L^3)$, and
$\xi'={\cal O}(L_s^4/L^3)$.
At the matching point,
$\bar\phi={\cal O}(V^4\kappa^3/\rho_c^3L^6)$, $\bar E={\cal O}(E_0)$, and
$\bar\rho={\cal O}(\rho_c)$.
In between these two subregimes,
when $L_s^4/L^3L_c$ is of order unity,
$\xi'$ vanishes and $\bar\phi_{\rm asym}(x)$ becomes an exact solution of
Eq.~(\ref{eq:finitet}) which also fulfills all boundary conditions.
In the same range, $\xi$ changes sign from positive to negative
values.

We conclude that the width of the boundary layer is of order
${\rm max}\,(L_c,L_s^4/L^3)$. At the matching point, $\bar E\ll V/L$.
The boundary condition (\ref{eq:bc1a}), used to calculate the shot-noise power $P$,
ignores the boundary layer. This is justified because $P$ is a bulk property.
We estimate the contribution to $P/P_{\rm Poisson}$ coming from the boundary layer 
to be of order ${\rm max}\,(L_c/L,(L_s/L)^4)$ (possibly to some positive power),
hence to be $\ll 1$ in the regime of space-charge limited conduction.

\section{Energy-dependent scattering time}
\label{sec:dep}

We consider now an energy-dependent scattering time.
We restrict ourselves to $d=3$ and assume a power-law dependence
$\tau(\varepsilon)=\tau_0\varepsilon^\alpha$. The energy-dependence
of the rate $1/\tau$ is governed by the product of the
scattering cross-section and the density of states.
For short-range impurity scattering the cross-section
is energy-independent, hence $\alpha=-1/2$.
This applies to uncharged impurities in semiconductors.
For scattering by a Coulomb potential the
cross-section is $\propto \varepsilon^{-2}$, hence
$\alpha=3/2$.
This applies to scattering by charged impurities in semiconductors
\cite{Chattopadhyay}.
The case $\alpha=0$ considered so far lies between these
two extremes \cite{nagaevnote}.
We have found an exact analytical solution for the case of short-range
scattering, to be presented below.
The case
of long-range impurity scattering remains an open problem, as
discussed at the end of this section.

For short-range impurity scattering, the
technical steps are similar to those of Sec.~\ref{sec:indep}.
We first determine the mean potential $\bar\phi(x)$. The scaling
properties of Eq.~(\ref{eq:dgl}) are exploited by introducing the
rescaled potential $\chi(x)$, with 
\begin{equation}
\bar\phi(x)=-\left(\frac{3e^{1/2}L^3\bar I}{2\mu_0\kappa A}\right)^{2/3}\chi(x/L)
.
\end{equation}
In this way we eliminate the dependence on the current $\bar I$ and the
length of the conductor $L$. The rescaled potential fulfills the
differential equation
\begin{equation}
\frac{1}{2}\chi^{-1/2}\frac{{\rm d}\chi}{{\rm d}x}
\frac{{\rm d}^2\chi}{{\rm d}x^2}
-
\chi^{1/2}
\frac{{\rm d}^3\chi}{{\rm d}x^3}
=1
,
\label{eq:meandep}
\end{equation}
with boundary conditions $\chi(0)=0$, $\chi'(0)=0$, $\chi''(1)=0$.

We substitute 
\begin{equation}
\chi(x)=\sum_{l=0}^\infty g_l x^{\eta l+2}
\end{equation}
into Eq.\ (\ref{eq:meandep}).
Power matching gives 
in the  first order $g_0=2^{-2/3}$. The second order leaves $g_1$ as a
free coefficient, but fixes the power $\eta=(\sqrt {13}  -1)/2$. The
coefficients
$g_l$ for $l\ge 2$ are then determined recursively as a function of
$g_1$. From the condition $\chi''(1)=0$ we obtain $g_1=-0.1808$.
The resulting series expansion converges rapidly, with the coefficient 
$g_{12}$ already of order $10^{-12}$.

\begin{figure}
\epsfxsize\columnwidth
\epsffile{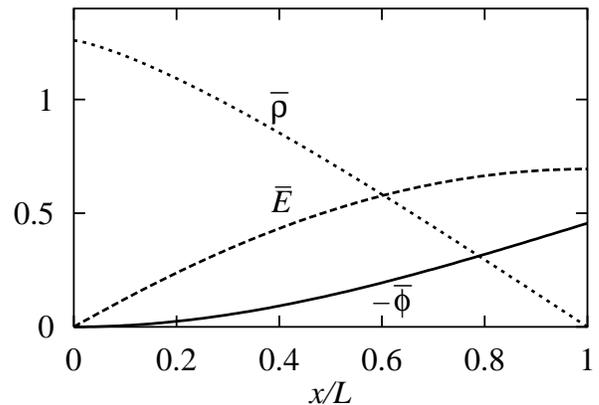}
\caption{
\narrowtext
Profile of the mean electrical potential $\bar \phi$ [in units of
$L^2(3e^{1/2}\bar I /2\mu_0\kappa A)^{2/3}$], 
the electric field $\bar E$ [in units of
$L (3e^{1/2}\bar I /2\mu_0\kappa A)^{2/3}$],
and the charge density $\bar \rho$
[in units of $(3e^{1/2}\bar I \kappa^{1/2}/2\mu_0 A)^{2/3}$] for
a three-dimensional conductor with short-range impurity scattering,
computed from Eq.~(\protect\ref{eq:meandep}).
}
\label{fig:fig6}
\end{figure}

The averaged potential and its first and second derivative are plotted
in Fig.~\ref{fig:fig6}. The electric field 
$\propto \chi'(x)$ increases now linearly
at the current source, hence the charge density $\propto \chi''(x)$ remains
finite there.
The current-voltage characteristic is
\begin{equation}
\bar I=
\frac{2\mu_0\kappa A}{3e^{1/2}L^3}\left(\frac{V}{\chi(1)}\right)^{3/2}
,
\end{equation}
with $\chi(1)=0.4559$.
This is a slower increase of $I$ with $V$ than the quadratic increase
(\ref{eq:iv}) in systems with energy-independent scattering.  

The rescaled fluctuations $\psi(x,t)$, introduced by
\begin{equation}
\delta
\phi(x,t)=-\left(\frac{3e^{1/2}L^3\bar I}{2\mu_0\kappa A}\right)^{2/3}\psi(x/L,t)
,
\end{equation}
fulfill the linear differential equation
\begin{eqnarray}
{\cal L}[\psi]&=&-\chi^{1/2}\frac{\partial^3 \psi}{\partial x^3}
+\frac 12\frac{\chi'}{\chi^{1/2}}
\frac{\partial^2 \psi}{\partial x^2}
+\frac 12 \frac{\chi''}{\chi^{1/2}}\frac{\partial \psi}{\partial x}
\nonumber\\
&&\qquad -\frac14
\left(
\frac{\chi'\chi''}{\chi^{3/2}}+2
\frac{\chi'''}{\chi^{1/2}}
\right)\psi
\nonumber
\\
&=&
\frac{\delta I(t)-\delta J(x,t)}{\bar I}
.
\end{eqnarray}
The solution of the inhomogeneous equation
is found with help of the three independent 
solutions
of the homogeneous equation ${\cal L}[\psi]=0$,
\begin{equation}
\psi_1(x)=\frac{\rm d}{{\rm d}x}\chi(x),
\end{equation}
\begin{equation}
\psi_2(x)=\chi(x)-\frac x2 \frac{\rm d}{{\rm d}x}\chi(x),
\end{equation}
\begin{eqnarray}
\psi_3(x)&=&
\psi_1(x)\int_x^1{\rm d}x'\,\frac{\chi^{1/2}(x')\psi_2(x')}{
{\cal W}^2(x')}
\nonumber
\\
&&{}
-
\psi_2(x)\int_x^1{\rm d}x'\,\frac{\chi^{1/2}(x')\psi_1(x')}{
{\cal W}^2(x')}
,
\end{eqnarray}
where we have defined
\begin{equation}
{\cal W}(x)=\psi_1(x)\psi_2'(x)- \psi_1'(x)\psi_2(x)
.
\end{equation}
The special solution which fulfills $\psi(0,t)=\psi'(0,t)=\psi(1,t)=0$
is
\begin{eqnarray}
\psi(x,t)&=&
\int_0^1{\rm d}x'\,\frac{\chi^{1/2}(x')}{{\cal W}^2(x')}
\Big[\Theta(x-x')\psi_1(x)\psi_2(x')
\nonumber
\\
&&
{}
+\Theta(x'-x)\psi_1(x')\psi_2(x)
-\frac{\psi_1(1)}{\psi_2(1)}\psi_2(x)\psi_2(x')
\Big]
\nonumber
\\
&&
{}
\times
\int_0^{x'}{\rm d}x''\,
\frac{\delta I(t)-\delta J(x'',t)}{\bar I }\frac{{\cal W}(x'')}{\chi(x'')}
.
\end{eqnarray}

The condition $\psi''(1,t)=0$ relates the fluctuating current $\delta I$
to the Langevin current $\delta J$. The resulting
expression is again of the form
(\ref{eq:deltai}), with now
\begin{equation}
{\cal C}=\int_0^1{\rm d}x\,{\cal G}(x),
\label{eq:calcdep}
\end{equation}
\begin{equation}
{\cal G}(x)=\frac{{\cal W}(x)}{\chi(x)}
\left(
2+\frac{{\chi'}^2(1)}{\psi_2(1)}\int_x^1{\rm d}x'\,
\frac{\chi^{1/2}(x')\psi_2(x')}{{\cal W}^2(x')}
\right)
.
\label{eq:calgdep}
\end{equation}
The shot-noise power is given by Eq.~(\ref{eq:pres1})
with ${\cal H}(x)$ as defined in Eq.~(\ref{eq:calh})
and the mean occupation number $\bar{\cal F}$ still given by 
Eq.\ (\ref{eq:barf}).
Instead of Eq.~(\ref{eq:calh2}) we now have
\begin{eqnarray}
{\cal H}(x)&=&2e\bar I\chi(x)\int_x^1{\rm d}x'\,\frac{1}{\chi(x')}
\nonumber
\\
&=&
P_{\rm Poisson}\chi^{1/2}(x)\chi''(x),
\label{eq:calhdep}
\end{eqnarray}
where we integrated with help of Eq.~(\ref{eq:meandep}) and used
$\chi''(1)=0$.

Collecting results we obtain the shot-noise suppression factor
\begin{equation}
P/P_{\rm Poisson}=0.3777,
\label{eq:depres}
\end{equation}
which is about $20\%$ larger than the result obtained in Sec.~\ref{sec:indep}
for an energy-independent scattering time in three dimensions.
Eq.\ (\ref{eq:depres}) can be compared with the $\alpha$-dependent
result in the drift approximation,
\begin{equation}
P/P_{\rm Poisson}=\frac{6(\alpha-1)(\alpha+2)(16\alpha^2+36\alpha-157)}{
5(2\alpha-5)(8\alpha-17)(13+8\alpha)}
 .
 \label{eq:driftapp}
\end{equation}
For $\alpha=-1/2$ the drift approximation
gives $P=0.4071 P_{\rm Poisson}$, about $10\%$ larger than the exact result
(\ref{eq:depres}). 

We now turn briefly to the case of long-range impurity scattering.
The kinetic equation (\ref{eq:dgl}), on which our analysis is based,
predicts a logarithmically diverging electric field $\propto -\ln^{1/3}x$ 
at the current source for $\alpha=1$.
In the range $\alpha> 1$, which includes the case $\alpha=3/2$ of scattering by
charged impurities, 
we could not determine the low-$x$ behavior.
[A behavior $\phi\propto Cx^\beta$ is ruled out because Eq.\ (\ref{eq:dgl})
cannot be satisfied with a real coefficient $C$.]
In the drift approximation, the shot-noise power (\ref{eq:driftapp})
vanishes as $\alpha\to 1$. Presumably, a non-zero answer for $P$ 
would follow for $\alpha\geq 1$ if the non-zero thermal energy and finite
charge density at the current source is accounted for. This remains
an open problem.

\section{Discussion}
\label{sec:concl}

We have computed the shot-noise power in a non-degenerate
diffusive semiconductor,
in the regime of space-charge limited conduction,
for two types of elastic impurity scattering.
In three-dimensional systems
the shot-noise suppression factor $P/P_{\rm Poisson}$
is close to $1/3$ both
for the case of an energy-independent scattering rate
($P/P_{\rm Poisson}=0.3097$) and for the case of short-range scattering
by uncharged impurities ($P/P_{\rm Poisson}=0.3777$).
(The latter case also applies to quasi-elastic scattering by acoustic phonons,
discussed below.)
Our results are in good agreement
with the numerical simulations for
energy-independent scattering by
Gonz{\'a}lez {\it et al.} \cite{Gonzalez}.
The results in the drift approximation \cite{Beenakker1998}
are about $10\%$ larger.
We found that capacitance fluctuations are strongly suppressed by the long-range
Coulomb interaction.
We discussed the effects of a non-zero thermal excitation
energy and a finite carrier density in the current source and determined
the regime  $L\gg L_s\gg L_c$ for
space-charge limited conduction
($L_s$ and $L_c$ being the screening lengths
in the semiconductor and current source, respectively).
Two subregimes of overscreening
and underscreening were identified,
again in quantitative agreement with the numerical simulations
\cite{Gonzalez}.

Let us discuss the conditions for experimental observability.
We have neglected
inelastic scattering events. These drive the gas of charge carriers towards
local thermal equilibrium and result
in a suppression of the shot noise down to thermal noise,
$P=8kT{\rm d}{\bar I}/{\rm d}V$ \cite{Beenakker1998}.
Inelastic
scattering by optical phonons can be neglected for voltages
$V< kT_D/e$, with $T_D$ the Debye temperature.
Scattering by acoustic phonons is quasi-elastic as long as the
sound velocity $v_s$ is much smaller than the typical electron velocity
$v\approx(eV/m)^{1/2}$.
For large enough temperatures $T\gg mvv_s/k$
the elastic scattering time 
$\tau \propto \epsilon^{-1/2}$ depends on energy in the same way as for
short-range impurity scattering \cite{Gantmakher}.

All requirements appear to be realistic for a semiconducting
sample with a sufficiently low carrier density:
The electron gas is degenerate even at quite low temperatures (a few Kelvin).
Short-range
electron-electron scattering is rare due to the diluteness of the
carriers. Scattering by phonons is predominantly elastic.
If the dopant (charged impurities) is sufficiently dilute, 
the impurity scattering is predominantly short-ranged.
Under these conditions we would expect the shot-noise power to be about
$1/3$ of the Poisson value.

\acknowledgements

We are indebted to J.~M.~J.~van Leeuwen 
for showing us how to
solve the non-linear differential equation 
(\ref{eq:mean}).
We thank T. Gonz{\'a}lez for permission to use the
data shown in Fig.~\ref{fig:fig5}.
Discussions with O. M. Bulashenko,
A. V. Khaetskii, and W. van Saarloos are gratefully
acknowledged.
This work was supported by the European Community
(Program for the Training and Mobility of Researchers)
and by the Dutch Science Foundation NWO/FOM.

\ecols

\begin{thebibliography}{10}
\bibitem{flatland}
A physical realization of the case $d=2$ would be a layered material in
which each layer
contains a two-dimensional electron gas.
A single layer would not suffice because the Poisson equation would then
remain three-dimensional and can not be reduced to the form 
(\ref{eq:poisson2}).
\bibitem{ineldrift}
In the case of inelastic scattering  the drift term $\mu\rho E$
follows, regardless of the energy dependence of the scattering rate,
from the
equilibrium condition
${\cal F}=-kT\partial {\cal F}/\partial \varepsilon$.
This condition does not hold when all scattering is elastic.
\end{thebibliography}
\end{document}